\pdfoutput=1
\documentclass[sigconf,screen]{acmart}

\usepackage{booktabs}
\usepackage{multirow}
\usepackage{graphicx}
\usepackage{makecell}
\usepackage{enumitem}
\usepackage{pifont}
\usepackage{tikz}
\usetikzlibrary{arrows.meta,positioning,fit,backgrounds,calc}
\usepackage[most]{tcolorbox}
\usepackage{listings}
\usepackage{wrapfig}

\newcommand{\bench}{\textsc{CASE-Bench}}

\newcommand{\SP}{\textsc{SecPass}}
\newcommand{\FP}{\textsc{FuncPass}}
\newcommand{\CbE}{\textsc{CbE}}

\newcommand{\cmark}{\ding{51}}
\newcommand{\xmark}{\ding{55}}

\newtcolorbox{finding}[1]{enhanced,breakable,colback=gray!6,colframe=gray!55,
  boxrule=0.5pt,arc=1pt,left=4pt,right=4pt,top=2pt,bottom=2pt,
  fonttitle=\small\bfseries,coltitle=black,colbacktitle=gray!18,title={#1}}

\begin{document}

\title[A.S.E 2.0]{A.S.E 2.0: Understanding How Repository Context\texorpdfstring{\\}{} Impacts the Security of LLM-Generated Code}

\author{Jiongchi Yu}
\orcid{0000-0002-2888-4499}
\authornote{Both authors contributed equally to this research.}
\email{jiongchiyu@acm.org}
\affiliation{%
  \institution{Nanyang Technological University}
  \city{Singapore}
  \country{Singapore}}

\author{Keke Lian}
\orcid{0009-0005-9026-8995}
\authornotemark[1]
\email{kekelian@tencent.com}
\affiliation{%
  \institution{Tencent}
  \country{China}}

\author{Junqing Wang}
\orcid{0009-0001-8457-5789}
\email{wangjunqing388@gmail.com}
\affiliation{%
  \institution{Zhejiang University}
  \city{Hangzhou}
  \country{China}}

\author{Zhikang Fan}
\orcid{0009-0006-5940-2807}
\email{15030857023@163.com}
\affiliation{%
  \institution{Zhejiang University}
  \city{Hangzhou}
  \country{China}}

\author{Zhaoxuan Li}
\orcid{0000-0002-2195-0799}
\email{lizhaoxuan@iie.ac.cn}
\affiliation{%
  \institution{IIE, CAS}
  \city{Beijing}
  \country{China}}

\author{Yongheng Huang}
\orcid{0009-0001-7874-6430}
\email{yonghenghuang384@gmail.com}
\affiliation{%
  \institution{Tencent}
  \country{China}}

\author{Ziming Zhao}
\orcid{0000-0003-1455-4330}
\authornote{Corresponding authors.}
\email{zhaoziming@zju.edu.cn}
\affiliation{%
  \institution{Zhejiang University}
  \city{Hangzhou}
  \country{China}}

\author{Dong Zhang}
\orcid{0009-0000-9706-8367}
\authornotemark[2]
\email{zalezhang@tencent.com}
\affiliation{%
  \institution{Tencent}
  \country{China}}

\renewcommand{\shortauthors}{Yu et al.}

\begin{abstract}
Large language model (LLM) coding assistants and agents construct a repository context before generating code.
For security-sensitive code, this context decides whether the model sees the project's safeguards (e.g., authorization checks) or insecure code to imitate.
However, context construction is designed almost exclusively for functional correctness, and secure-code benchmarks treat the context as fixed, so how repository context affects the security of generated code remains unclear.
We construct \bench{}, a benchmark of 145 repository-level tasks from real CVEs in 51 projects.
Each task runs in Docker, has functional tests and a proof-of-concept exploit validated in both directions, and is annotated with the repository code that enforces or undermines its security.
Using \bench{}, we compare 13 context construction methods and six controls with six LLMs, edit single context elements, and analyze coding-agent trajectories.
We find that \ding{182}~functionality-oriented context raises the functional pass rate by up to 8.2 percentage points (pp) but the secure pass rate by at most 3.7 pp, \ding{183}~withholding a safeguard lowers the secure pass rate by up to 8.1 pp, while one inserted insecure exemplar raises correct-but-exploitable code by 6.3 pp, and \ding{184}~coding agents observe the relevant safeguard in only \mbox{27-36\%} of trajectories.
These findings inform security-aware context construction and secure-code evaluation.

\end{abstract}




\maketitle

\section{Introduction}
\label{sec:introduction}

LLM coding assistants and agents rarely generate code from the prompt alone.
They first assemble \emph{context} from the repository (e.g., similar code or imported definitions)~\cite{zhang2023repocoder,liu2024repobench,yang2024sweagent,wang2025openhands}.
For security-sensitive code, the information that makes an implementation safe is often not local: a parser may need the zero-divisor check of a sibling routine, and an endpoint may need a project-specific permission.
Context construction therefore decides whether the model sees the repository's security policy, and it can also expose legacy code with the same weakness (\S\ref{subsec:motivating-example}).

Context construction has nevertheless been designed and evaluated almost exclusively for functional correctness~\cite{zhang2023repocoder,wang2025coderagbench,gu2025whattoretrieve,hai2025repoexec}.
Secure-code benchmarks have moved from snippets~\cite{siddiq2022securityeval,bhatt2023cyberseceval,peng2025cweval} to real repositories~\cite{lian2026ase,shen2025secrepobench,chen2026securevibebench,zhao2026susvibes,wang2026realsecbench}, and RealSec-bench~\cite{wang2026realsecbench} observes that retrieval improves correctness more than security.
However, existing benchmarks cannot explain \emph{how} context affects security, because most of them share four limitations.
\ding{182}~\emph{Context is fixed, not studied.}
They evaluate one context configuration or compare a few retrievers, so no benchmark explains which context properties cause the gap between correctness and security.
\ding{183}~\emph{Security oracles are static or untargeted.}
Many benchmarks judge security with static rules or LLM judges~\cite{lian2026ase,hajipour2024codelmsec,wang2026realsecbench}, which miss runtime-dependent weaknesses and cannot show that the specific weakness is exploitable.
\ding{184}~\emph{Functionality is not validated jointly.}
Without functional tests, broken or deny-all code can look secure~\cite{lian2026ase,siddiq2022securityeval}.
\ding{185}~\emph{Security-relevant context is not measurable.}
No benchmark records which repository code enforces or undermines security, so outcomes cannot be attributed to the content of a context.

To address these limitations, we construct \bench{}, a benchmark of 145 repository-level generation tasks derived from 144 real CVEs in 51 open-source projects.
The tasks are \emph{diverse}, covering 28 CWEs in seven languages, and they span \emph{different difficulty levels}, from generic checks to project-specific policies.
They are also of \emph{high quality}: every task is reproducible in a Docker environment, and its functional tests and proof-of-concept (PoC) exploit are validated against both the vulnerable and the patched version (\ding{183}, \ding{184}).
Two experts annotated the security-relevant repository code of every task (\ding{185}).
On \bench{}, we hold the task, prompt, and oracle fixed and vary only the context (\ding{182}).
We compare 13 context construction methods and six controls with six LLMs, edit single context elements, and analyze what coding agents observe.
Our study answers the following research questions:
\begin{itemize}[leftmargin=1.2em,itemsep=2pt,topsep=3pt]
\item \textbf{RQ1: How do context construction methods affect the correctness and security of generated code?} We measure the functional and secure pass rates of real methods against the target file alone and against security controls.
\item \textbf{RQ2: Which context properties make context help or hurt security?} We edit single context elements, such as safeguards or insecure exemplars, and analyze why code remains insecure.
\item \textbf{RQ3: Do coding agents gather security-relevant context?} We analyze what agents search for and observe, and how this relates to the security of their code.
\end{itemize}
We additionally check whether the conclusions depend on the security oracle.

Our study shows that functionality-oriented context raises the functional pass rate by up to 8.2 percentage points (pp) but the secure pass rate by at most 3.7 pp, with no significant difference from the target file alone.
Context helps when it exposes concrete safeguards: withholding a safeguard lowers the secure pass rate by up to 8.1 pp, whereas a security-themed placebo has almost no effect.
It hurts when it exposes insecure exemplars, one of which raises correct-but-exploitable code by 6.3 pp even with the safeguard present.
Coding agents share this bias toward functionality, observing the relevant safeguard in only 27-36\% of their trajectories.

\textbf{Contributions.}
In summary, we make three contributions:
\begin{itemize}[leftmargin=1.2em,itemsep=2pt,topsep=3pt]
\item We construct \bench{}, an executable benchmark whose tasks are validated in both directions, solvable within the edit scope, and annotated with security-relevant repository code.
\item We conduct a controlled study of 13 context construction methods and six controls with six LLMs (49,590 generations), which extends the functionality-security gap reported for a few retrievers to all major method families.
\item Through single-element context edits, failure patterns, and 1,305 agent trajectories, we show which elements change security and derive guidelines for security-aware context construction.
\end{itemize}

\section{Problem and Motivating Example}
\label{sec:background}

\subsection{Problem Formulation}
\label{subsec:formulation}

\textbf{Task and context.}
A task comprises a repository at the last vulnerable commit, a target file with a removed code region (the \emph{masked region}), and a requirement that specifies its functionality.
The model must regenerate the masked region and is never told that the task is security-relevant (\S\ref{subsec:construction}).
A context construction method selects repository units (e.g., functions) to accompany the requirement and the masked target file.

\textbf{Outcomes.}
Each generated implementation is built, tested, and attacked with the task's exploit, and it receives one of four outcomes:
\emph{Unbuildable}, \emph{Incorrect} (a functional test fails), \emph{correct but exploitable} (\CbE{}: all tests pass, but the exploit succeeds), or \emph{Secure} (all tests pass, and the exploit shows no effect).
Both labels are relative to the task's oracles: ``correct'' means passing the provided tests, and a failed exploit does not rule out other vulnerabilities (\S\ref{subsec:oracles}).

\textbf{Context effect.}
Let $\textit{SecRate}(s,t)$ be the fraction of runs in which method $s$ yields a secure implementation of task $t$.
The context effect of $s$ on $t$ is $\Delta(s,t) = \textit{SecRate}(s,t) - \textit{SecRate}(\text{Local-only},t)$, where Local-only shows the model only the target file.
All else fixed, systematic differences in $\Delta$ are attributable to the context, while differences over three runs also contain sampling noise (\S\ref{subsec:statistical-analysis}).

\subsection{Motivating Example}
\label{subsec:motivating-example}

\begin{figure*}[t]
\centering
\includegraphics[width=0.85\textwidth]{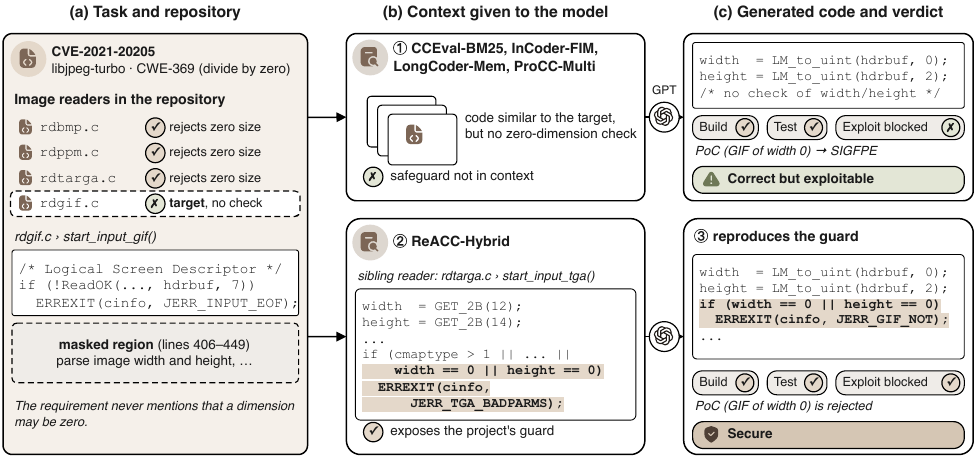}
\caption{CVE-2021-20205. (a) Sibling image readers check for zero dimensions; the GIF reader does not. (b, c) With the model, task, and prompt fixed, \ding{172}~four contexts yield correct-but-exploitable code, whereas \ding{173}~ReACC-Hybrid retrieves the Targa guard and \ding{174}~the generated code reproduces it. Outcomes hold across all runs.}
\Description{A GIF parser example contrasts generated code without a dimension check with code that reproduces a guard retrieved from a sibling image reader.}
\label{fig:motivating-example}
\end{figure*}

Figure~\ref{fig:motivating-example} shows a task derived from CVE-2021-20205, a divide-by-zero vulnerability in the GIF reader of libjpeg-turbo.
The requirement asks for parsing the image dimensions but does not mention that they may be zero, whereas the other image readers of the same repository reject zero dimensions.
With four functionality-oriented contexts that lacked such a check, the code of GPT-5.4 passed all tests but was exploited by a zero-width GIF in every run.
With the ReACC-Hybrid context, which contained the guard of the Targa reader, the model reproduced the guard and was secure in every run.
Similarity retrieval exposed the guard here, but it could equally have retrieved an unchecked sibling; our controlled context edits separate these effects.

\subsection{How Context Can Help or Hurt}
\label{subsec:hypotheses}

As an analysis framework, we relate each context unit to the \emph{security policy} that the masked region must enforce (e.g., a bounds check or an authorization rule).
A unit may serve several roles, and some units remain unclassified.
From prior work, we derive four candidate mechanisms, which we treat as hypotheses.

\textbf{Exposure.} A unit that enforces the policy (a \emph{safeguard}) can help: retrieved APIs improve correctness~\cite{hai2025repoexec,gu2025whattoretrieve}, and retrieved secure examples improve function-level security~\cite{zhang2024seccoder,lin2025securitycourse,shi2026rescue}.

\textbf{Imitation.} A unit that violates the policy (an \emph{insecure exemplar}, e.g., an unfixed sibling) can hurt: surrounding code changes the security of suggestions~\cite{pearce2022asleep}, and planted vulnerable examples induce vulnerable code~\cite{lin2025poisoning}.

\textbf{Dilution.} Unrelated units and long contexts can bury a safeguard~\cite{shi2023distracted,liu2024lost,du2025contextlength}.

\textbf{Priming.} Security-themed text that does not apply to the target may change the output by itself~\cite{tony2025prompting}, which confounds the evaluation of security-aware methods.

\section{\bench{}}
\label{sec:benchmark}

\begin{figure*}[t]
\centering
\includegraphics[width=0.85\textwidth]{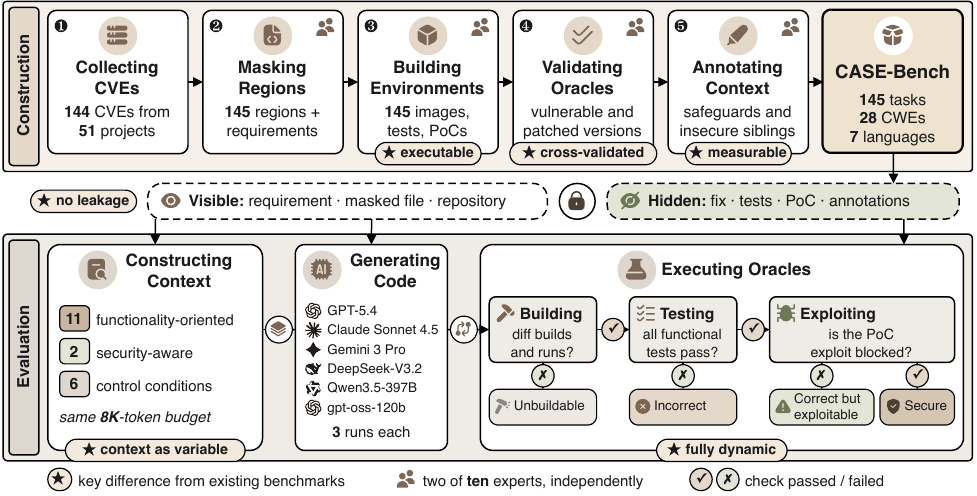}
\caption{Overview of \bench{}: construction (top), the model--evaluation information barrier (middle), and controlled evaluation (bottom). $\bigstar$ marks properties compared with prior benchmarks in \S\ref{sec:related-work}.}
\Description{The benchmark workflow proceeds from CVE selection through task construction, executable validation, context annotation, and controlled evaluation. A barrier separates model-visible inputs from evaluation artifacts.}
\label{fig:overview}
\end{figure*}

\subsection{Task Construction and Validity}
\label{subsec:construction}

\bench{} was built by ten security experts (five Ph.D.\ and five M.S.\ students with at least two years of experience in vulnerability analysis) in five steps (Figure~\ref{fig:overview}); every manual step was performed independently by two experts.

\textbf{Selection.}
We collected reproducible CVEs (e.g., with a public PoC) and prioritized the CWE Top~25 and OWASP Top~10. We excluded fixes that change more than ten files, because they cannot be localized to one coherent region; 488 of 1,236 records passed this screening.
Because rebuilding a historical environment takes several person-hours, we selected 236 candidates by two criteria.
\emph{Reproducibility} requires that the base commit builds with an available toolchain and that the PoC replays in an isolated container.
\emph{Diversity} balances weakness types and code bases while limiting near-duplicate CVEs.
This purposive sample favours reproducible CVEs, and it used only CVE metadata, never model outputs.
We admitted 144 CVEs (145 tasks); most rejections stemmed from build failures of old commits (41), PoCs that did not reproduce (30), or vulnerable code that spanned no coherent region (21).

\textbf{Task definition.}
An expert identified the base commit (the parent of the fix) and the vulnerable region that the fix changes, which becomes the masked region (1 to 1,207 lines, median 10).
The expert then wrote a requirement that specifies the intended behaviour (e.g., inputs and intended users) but names no attack or defense.
Requirements were drafted with an LLM; two experts audited all of them and rewrote 23.
Requirements do not restate policies defined in the visible repository, and the derivability annotation (\S\ref{subsec:annotation}) records whether the security property can be inferred from the available information.
In the 51 \emph{multi-location} tasks, the fix also changes other files, which stay at the vulnerable version.

\textbf{Information barrier.}
The model receives the requirement, the target file with a placeholder for the region, and the context, and it returns a unified diff.
It never sees vulnerability information (e.g., the CVE identifier or the fix) or our evaluation artifacts (e.g., the tests).
Context methods retrieve only from the \emph{visible repository}, i.e., the base snapshot stripped of the masked region and of any trace of the vulnerability (e.g., the commit history).
Because the snapshot precedes the fix, it may contain \emph{insecure exemplars} (e.g., unfixed siblings with the same weakness). We keep and annotate them, as real repositories contain such code (74 tasks).

\textbf{Contamination.}
We flag near-copies of the vulnerable or patched region (edit similarity $\geq 0.9$), build semantics-preserving variants of 40 tasks~\cite{xia2024evoeval}, and report 19 tasks from 2024 or later separately.

\subsection{Executable Evaluation}
\label{subsec:oracles}

Each generation is evaluated in a fresh container in three stages.
\textbf{Build:} the diff applies, and the project builds and starts (timeout 5\,min).
\textbf{Functional tests:} all expert-written tests pass (4.6 inputs per task on average; timeout 2\,min). The tests cover typical and boundary inputs (e.g., benign inputs along the PoC path), so deny-all code fails them; correctness remains relative to these tests.
\textbf{Exploit:} the PoC is replayed (timeout 2\,min), and exploitation is decided by an observable effect, e.g., a sanitizer report for memory-safety weaknesses~\cite{lee2025secbench,mei2024arvo} or an unauthorized action for web weaknesses, so secure implementations that differ from the historical fix are not penalized~\cite{lee2026secbenchpro}.
Each PoC defines its effect check in advance, and \emph{Secure} requires that the PoC completes without that effect; infrastructure failures trigger re-execution rather than a verdict.
For 40 tasks sampled by CWE family, we wrote three exploit variants each (e.g., encoded payloads).
For comparison with static evaluation (\S\ref{subsec:measurement}), we also run CWE-specific CodeQL and Joern queries~\cite{codeql,joern} and count only alarms whose path crosses the regenerated region.

\begin{table*}[t]
\centering
\normalsize
\caption{Oracle validation (all 145 tasks unless stated otherwise).}
\label{tab:oracle-validation}
\begin{tabular}{llr}
\toprule
\textbf{Check} & \textbf{Criterion} & \textbf{Result} \\
\midrule
V1 Vulnerable version & builds, tests pass, PoC succeeds & 145/145 \\
V2 Patched version & builds, tests pass, PoC fails & 145/145 \\
V3 Alternative secure implementation & builds, tests pass, PoC fails & 143/145 \\
V4 Deny-all implementation & tests fail & 145/145$^{a}$ \\
V5 Determinism & identical verdicts & 145/145$^{b}$ \\
V6 Manual audit & agreement / Cohen's $\kappa$~\cite{cohen1960kappa} & 95.7\% / 0.90 \\
\bottomrule
\multicolumn{3}{l}{\footnotesize $^{a}$after strengthening the tests of 6 tasks; $^{b}$after stabilizing the start-up scripts of 3 tasks.}
\end{tabular}
\end{table*}

\textbf{Oracle validation.}
Table~\ref{tab:oracle-validation} summarizes the validation.
An independent reviewer first wrote the alternative implementation (V3) and then reviewed and replayed the task on both versions (V1, V2); 38 tasks needed a second review round, and 9 of them a third.
V3 has two purposes.
It checks that the oracles do not overfit the historical fix, because the reviewer wrote the alternative before seeing the patch.
It also checks solvability within the edit scope, because the alternative replaces only the masked region while all other files stay vulnerable, which matters for multi-location tasks.
The two tasks that fail V3 remain solvable within this scope, since the patched region alone passes all checks.
V1 shows that the tests do not encode the security property, V4 that deny-all code cannot score as secure, and V6 that the verdicts agree with manual exploitation.

\subsection{Context Annotations and Dataset Profile}
\label{subsec:annotation}

Two experts independently annotated every task, and a third resolved disagreements ($\kappa$ over each task's candidate units).

\textbf{Security roles} ($\kappa=0.82$).
\emph{Flow roles} (source and sink) locate the attack path, \emph{enforcement roles} (e.g., a validator or safe API) implement a check, and the \emph{policy} role defines what is allowed.
A \emph{safeguard} is a unit with an enforcement or policy role, and a task's \emph{acceptable safeguard sets} include the historical fix and annotated alternatives.
An \emph{insecure exemplar} processes the same untrusted input without a guard and is exploitable in the same way.
Safeguards and the source-sink path form the task's \emph{reference context}.

\textbf{Security locality} ($\kappa=0.79$).
It measures how much a secure solution needs the repository: generic built-ins suffice at \textbf{L1}, a project API is needed at \textbf{L2}, and a repository-specific rule decides at \textbf{L3}.

\textbf{Derivability} ($\kappa=0.77$).
It records whether the security property follows from the requirement and target file (38 tasks), from the reference context (99), or from neither (8).
We also extract \emph{fix-relevant APIs}, i.e., definitions that the patched region references; as they follow the historical fix, their recall is only a functional proxy.

\begin{figure*}[t]
\centering
\includegraphics[width=0.75\textwidth]{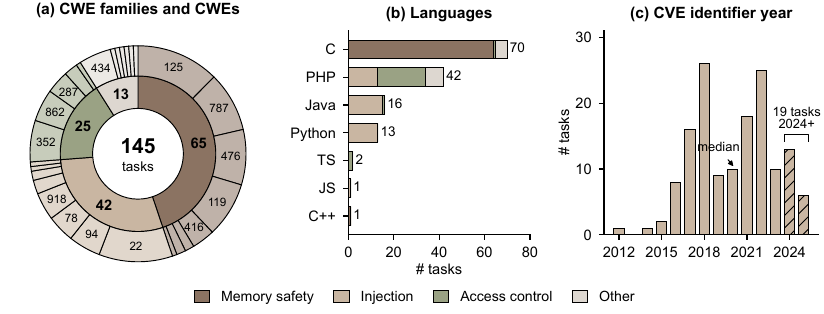}
\caption{Distribution of the 145 tasks. (a) CWE families (inner ring) and CWEs (outer ring). (b) Languages by family. (c) Year of the CVE identifier; hatched: 2024 or later.}
\Description{Three panels show the distribution of tasks across CWE families and individual CWEs, programming languages, and CVE years.}
\label{fig:benchmark-dist}
\end{figure*}

\textbf{Profile.}
\bench{} covers 28 CWEs, including 20 of the 2024 CWE Top~25, in four families led by memory safety (65 tasks) and injection (42) (Figure~\ref{fig:benchmark-dist}).
The tasks span seven languages and all three locality levels (41, 67, and 37 tasks at L1 to L3).
Locality differs by family: access-control tasks are almost all L3, whereas most memory-safety tasks are L1 or L2.
Six projects contribute 67 tasks, so we treat the project as the unit of clustering in all statistical analyses.
The median project has 86 KLOC (1.1M tokens), and only 6\% of projects fit into a 128K-token context.
The CVEs date from 2012 to 2025 (median 2020), and 19 tasks stem from 2024 or later, which reduces the risk that all tasks were seen during training.

\section{Study Design}
\label{sec:study-design}

\subsection{Context Conditions}
\label{subsec:conditions}

\textbf{Methods.}
We selected eleven \emph{functionality-oriented} methods that are published or deployed in widely used tools and that cover all major retrieval families (Table~\ref{tab:strategies}).
We compare their \emph{context policies} as reproduced in a common pipeline, not the original systems, and name each policy after its source (e.g., CCEval-BM25 for CrossCodeEval).
We add two \emph{security-aware} methods.
CodeGuarder~\cite{lin2025securitycourse} injects external security knowledge mined from vulnerability databases, rebuilt without the CVEs of \bench{}.
Source-sink slicing probes \emph{repository-internal} security context: following data-flow-guided retrieval~\cite{cheng2024draco}, it retrieves the path from untrusted input to a sensitive operation together with its guards.
It uses only the generic CodeQL and Joern models, never the oracle's queries, and proceeds in five steps.
\ding{182}~\emph{Anchoring:} sources are untrusted values that reach the inputs of the masked region, and sinks are sensitive operations reachable from the code that consumes its result.
\ding{183}~\emph{Candidates:} the analysis collects the paths between them within two call-graph hops, together with their guards, i.e., conditions that decide whether execution reaches the sink.
\ding{184}~\emph{Request semantics:} for web applications, paths link an endpoint to the forms that call it, and guards are recognized by the return values and state that decide whether a request proceeds (e.g., an early exit on a failed permission check), so project-specific wrappers need no hand-written models.
\ding{185}~\emph{Ordering:} shortest paths come first, each as its guards followed by its units in flow order, until the cap.
\ding{186}~\emph{Failure:} without a path, no units are returned, and the prompt equals Local-only.
For OpenEMR (\S\ref{subsec:case-analysis}), step~\ding{184} links the amendment request to its form and retrieves the guard \textit{aclCheckCore('patients', 'amendment', ...)}; the implementation and all slices are released.

\textbf{Common pipeline.}
Each method receives the requirement, the masked target file, and the visible repository, and it returns an ordered list of units.
We keep each method's ranking logic and hyperparameters and replace only language-specific parts (e.g., parsers by tree-sitter grammars), so for newly supported languages the policy is inspired by, not validated against, the original.
Every prompt contains the task inputs (e.g., the complete masked target file) and a context block, which is empty for Local-only.
The units are deduplicated among themselves, serialized in rank order, and truncated at a \emph{common retrieval cap} of 8K tokens that excludes the target file.
The cap equalizes neither length nor cost, so we report actual sizes (Table~\ref{tab:strategies}) and native budgets.
Deduplication does not remove code that repeats the target file, so the in-file methods differ from Local-only: InCoder-FIM re-presents the code around the region next to the generation point, and LongCoder-Mem adds imported declarations, which raises fix-API recall from .12 to .19.
Generator-independent contexts are cached and identical across models and runs.

\textbf{Controls.}
Six conditions anchor and de-confound the comparison.
\textbf{Local-only} shows the masked target file alone and is the reference for all context effects.
\textbf{BM25+README}, the configuration of A.S.E~\cite{lian2026ase}, anchors our results to a prior evaluation.
\textbf{Repo-128K} serializes the visible repository from the target's directory outward in the format of repository-level benchmarks~\cite{liu2024repobench,ding2023crosscodeeval} and truncates it so that the complete prompt and an output reserve fit into the smallest window among our models (118K tokens of content on average); as only 6\% of projects fit, it is a truncated repository, not the full one.
The \textbf{Reference context} contains exactly the annotated safeguards and source-sink path; it is a diagnostic, not an upper bound.
The \textbf{Security reminder} and the \textbf{Placebo} add a security instruction or security-themed code from another project, respectively; they bound rather than isolate priming, because foreign code may still convey generic checks.

\begin{table*}[t]
\centering
\footnotesize
\setlength{\tabcolsep}{2.4pt}
\caption{Context construction methods and conditions. ``Gen.-dep.'': the context depends on the generator model. ``Tokens'': mean context size beyond the masked target file, at the 8K retrieval cap (Repo-128K: 128K). ``Fix-API'': mean fraction of fix-relevant APIs whose definitions are in the prompt, including the masked target file (hence non-zero for Local-only).}
\label{tab:strategies}
\begin{tabular}{lllcrr}
\toprule
\textbf{Family} & \textbf{Method} & \textbf{Context policy (as reproduced)} & \makecell{\textbf{Gen.-}\\\textbf{dep.}} & \textbf{Tokens} & \textbf{Fix-API} \\
\midrule
\multirow{6}{*}{\makecell[l]{Reference/\\control}} & Local-only & masked target file only & - & 0 & .12 \\
 & BM25+README~\cite{lian2026ase} & BM25-ranked files + README & - & 7.6K & .29 \\
 & Repo-128K & serialized repository, truncated to 128K tokens & - & 118K & .63 \\
 & Security reminder & Local-only + generic security instruction & - & 0 & .12 \\
 & Placebo & Local-only + security code of another project & - & 3.9K & .12 \\
 & Reference context & annotated safeguards + source-sink path & - & 2.7K & 1.00 \\
\midrule
\multirow{2}{*}{In-file} & InCoder-FIM~\cite{fried2023incoder} & prefix and suffix around the masked region & - & 3.1K & .14 \\
 & LongCoder-Mem~\cite{guo2023longcoder} & long in-file window; imports/globals as memory & - & 6.8K & .19 \\
\midrule
\multirow{4}{*}{Similarity} & ReACC-Hybrid~\cite{lu2022reacc} & similar code, hybrid lexical + dense retrieval & - & 7.8K & .38 \\
 & CCEval-BM25~\cite{ding2023crosscodeeval} & cross-file chunks by BM25 on the in-file prefix & - & 7.9K & .27 \\
 & RepoBench-R~\cite{liu2024repobench} & imported cross-file definitions, ranked & - & 5.2K & .33 \\
 & ProCC-Multi~\cite{tan2024procc} & LLM-prompted retrievers chosen by a bandit & \cmark & 7.7K & .36 \\
\midrule
Iterative & RepoCoder-Iter~\cite{zhang2023repocoder} & retrieval re-queried with a draft completion & \cmark & 7.9K & .34 \\
\midrule
\multirow{2}{*}{Structure} & RLPG-Proposal~\cite{shrivastava2023rlpg} & prompt proposals (imports, siblings, parents) & - & 6.4K & .31 \\
 & DevEval-Dep~\cite{li2024deveval} & local file + extracted dependencies & - & 6.1K & .42 \\
\midrule
\multirow{2}{*}{Agentic} & SWE-agent-Ret~\cite{yang2024sweagent} & regions found by SWE-agent's search tools & - & 6.9K & .35 \\
 & ClaudeCode-Ret & regions selected by Claude Code (Read/Grep) & - & 7.2K & .47 \\
\midrule
\multirow{2}{*}{\makecell[l]{Security-\\aware}} & CodeGuarder~\cite{lin2025securitycourse} & retrieval + external security knowledge & - & 7.5K & .33 \\
 & Source-sink & CWE-agnostic source$\rightarrow$sink slice + guards & - & 4.1K & .74 \\
\bottomrule
\end{tabular}
\end{table*}

\subsection{Models, Generation, and Metrics}
\label{subsec:models}

We evaluate six generator LLMs from five developers, three proprietary (e.g., GPT-5.4) and three open-weight (e.g., Qwen3.5-397B-A17B).
Reasoning models use the providers' default settings, and the two non-reasoning configurations use temperature 0.2.
All conditions share one prompt skeleton and run three times per task.
RQ1 has 49,590 generations, RQ2 adds 49,140 with three models, and RQ3 adds 1,305 agent trajectories (total cost USD 38,000).

\textbf{Metrics.}
For a task set $T$ and $n=3$ runs per task, $\SP@1 = \frac{1}{|T|}\sum_{t \in T} \textit{secure}(t)/n$, where $\textit{secure}(t)$ counts the secure runs on $t$; this is the unbiased pass@1 estimator~\cite{chen2021codex} with ``secure'' as the success criterion~\cite{shen2025secrepobench}.
\FP@1 and \CbE@1 are defined analogously, so $\FP@1 = \SP@1 + \CbE@1$.
\SP@1 is our primary metric and \CbE@1 our primary risk metric; differences are reported in percentage points (pp).
Before generation, we characterize each prompt with context metrics (e.g., whether it covers an acceptable safeguard set).

\subsection{Interventions, Agents, and Failure Patterns}
\label{subsec:rq-designs}

\textbf{RQ2: context edits and failure patterns.}
We vary the retrieval cap from 1K to 32K tokens for four methods, and we apply single-element edits to two base contexts, the Reference context and the context of the best functionality-oriented method.
Each edit is compared with its unmodified base on all eligible tasks.
\textsc{Remove} and \textsc{Add} delete one safeguard role or insert the missing safeguards; because a removed safe API may also carry functional knowledge, we report \CbE@1 next to \SP@1.
\textsc{Insecure} inserts one naturally occurring insecure exemplar at a random position (74 tasks), and we never synthesize vulnerable code.
\textsc{Distract} appends irrelevant code of 2K to 32K tokens.
Two \emph{matched controls} replace a removed role, or the inserted exemplar, with irrelevant code of the same length at the same position, which separates content from length and position.
Finally, \textsc{Position} moves the safeguard within contexts of 4K to 64K tokens, and \textsc{Rename} gives it a neutral name repository-wide.

To explain why code remains insecure, each correct-but-exploitable generation receives one or more of five failure labels (Table~\ref{tab:rq5-failures}); for example, \textsc{Unused} means that a safeguard is present but neither called nor replicated, and \textsc{Imitated} requires an AST similarity of at least 0.7 to an exemplar in the context.
Equivalent checks are detected with the guard analyses of CodeQL and Joern~\cite{codeql,joern}.
Two authors labeled a stratified sample of 200 generations manually, and we also compare the dominant labels of two methods within each task-model pair, which avoids changes in the denominator.
The labels describe co-occurrence, not cause.

\textbf{RQ3: agents.}
We run SWE-agent~\cite{yang2024sweagent} with two backbones and Claude Code, three runs per task, in an isolated sandbox (e.g., without network access).
Agents may explore and run the whole repository but edit only the masked region (at most 50 steps and USD 2 per task).
This restriction keeps the task one of secure \emph{generation}, since edits elsewhere would turn it into repository-wide repair, and it keeps agents comparable with retrieval-based methods.
The \emph{observed context} of a trajectory is every region that a tool returned to the agent after output truncation.
Observed tokens accumulate over the trajectory, so they do not measure a single prompt, and we do not track whether a region remains in the agent's context at its final edit.
For each trajectory, we record whether the safeguard was observed before the first edit of the masked region and whether the run hit the step or cost cap.
Search queries are classified automatically as functional or security-related; identifiers (e.g., wrapper names) count as functional unless they contain a security term, and two authors validated this classification on 300 queries.

\textbf{Oracle comparison.}
We cross-tabulate static and exploit verdicts for all functionally correct RQ1 generations, calibrate the static queries on the known versions of each task, and compare method rankings on the same 145 tasks.

\subsection{Statistical Analysis}
\label{subsec:statistical-analysis}

Because tasks are clustered within projects~\cite{arcuri2011practical}, we report 95\% cluster-bootstrap intervals over the 51 projects~\cite{efron1994bootstrap} and test with logistic mixed-effects models~\cite{bates2015lme4} with random intercepts for project, task, and task-model pair.
We test 18 pre-specified comparisons with Holm correction~\cite{holm1979simple}: each of the 13 methods against Local-only, the two security-aware methods against both priming controls, and the Reference context against the best functionality-oriented method.
We regard differences of at least 3 pp as practically relevant, a threshold fixed before the analysis.
The best method is selected on the same data, which makes the Reference-versus-best difference conservative.
A power analysis on a pilot run showed 80\% power for 5.6 pp after correction, so a nonsignificant result does not establish the absence of an effect.
We compare rankings with Kendall's $\tau_b$~\cite{kendall1938} and report Cohen's $\kappa$~\cite{cohen1960kappa} for manual annotations.
All other analyses (e.g., robustness checks and agent analyses) are exploratory and reported as estimates; runs that fail for infrastructure reasons are re-executed, and outputs are never resampled based on their content.

\section{Results}
\label{sec:results}

%% ===================================================================
\subsection{RQ1: How Does Context Construction Affect Correctness and Security?}
\label{subsec:rq1}

\begin{table*}[t]
\centering
\small
\setlength{\tabcolsep}{4pt}
\caption{RQ1: Outcomes per method and condition (rates in \%, 6 models, 3 runs, 145 tasks). Context methods use the 8K retrieval cap. $\Delta$: change relative to Local-only in pp; bold: at least the 3~pp practical-relevance threshold; $\dagger$: Holm-adjusted $p<.05$ against Local-only. Best-of-methods selects, per task and model, the best functionality-oriented method on two runs and evaluates it on the third (cross-fitted, not deployable).}
\label{tab:rq1-main}
\begin{tabular}{lrrrrr}
\toprule
\textbf{Method / condition} & \textbf{\FP@1} & \textbf{\SP@1} [95\% CI] & \textbf{\CbE@1} & \textbf{$\Delta$\FP} & \textbf{$\Delta$\SP} \\
\midrule
\multicolumn{6}{l}{\textit{Reference / control}} \\
\quad Local-only & 56.9 & 12.6 [9.3, 16.2] & 44.3 & - & - \\
\quad BM25+README & 59.7 & 12.4 [9.1, 16.0] & 47.3 & +2.8 & -0.2 \\
\quad Security reminder & 55.1 & 13.9 [10.5, 17.6] & 41.2 & -1.8 & +1.3 \\
\quad Placebo & 56.2 & 12.9 [9.6, 16.5] & 43.3 & -0.7 & +0.3 \\
\quad Repo-128K & 60.9 & 12.1 [8.9, 15.6] & 48.8 & +4.0 & -0.5 \\
\midrule
\multicolumn{6}{l}{\textit{Functionality-oriented}} \\
\quad InCoder-FIM & 57.4 & 12.8 [9.5, 16.4] & 44.6 & +0.5 & +0.2 \\
\quad LongCoder-Mem & 57.8 & 13.2 [9.9, 16.8] & 44.6 & +0.9 & +0.6 \\
\quad ReACC-Hybrid & 61.9 & 15.4 [11.9, 19.2] & 46.5 & +5.0 & +2.8 \\
\quad CCEval-BM25 & 60.3 & 11.5 [8.3, 15.0] & 48.8 & +3.4 & -1.1 \\
\quad RepoBench-R & 52.1 & 11.8 [8.6, 15.3] & 40.3 & -4.8 & -0.8 \\
\quad ProCC-Multi & 65.1 & 14.9 [11.5, 18.6] & 50.2 & +8.2 & +2.3 \\
\quad RepoCoder-Iter & 55.0 & 12.9 [9.6, 16.5] & 42.1 & -1.9 & +0.3 \\
\quad RLPG-Proposal & 54.6 & 12.3 [9.0, 15.9] & 42.3 & -2.3 & -0.3 \\
\quad DevEval-Dep & 56.4 & 13.3 [10.0, 16.9] & 43.1 & -0.5 & +0.7 \\
\quad SWE-agent-Ret & 55.0 & 11.7 [8.5, 15.2] & 43.3 & -1.9 & -0.9 \\
\quad ClaudeCode-Ret & 63.8 & 16.3 [12.8, 20.1] & 47.5 & +6.9 & \textbf{+3.7} \\
\midrule
\multicolumn{6}{l}{\textit{Security-aware}} \\
\quad CodeGuarder & 58.8 & 16.8 [13.2, 20.7] & 42.0 & +1.9 & \textbf{+4.2} \\
\quad Source-sink & 61.2 & 18.1 [14.4, 22.1] & 43.1 & +4.3 & \textbf{+5.5}$^{\dagger}$ \\
\midrule
\multicolumn{6}{l}{\textit{Diagnostic (not deployable)}} \\
\quad Reference context & 62.3 & 24.8 [20.7, 29.2] & 37.5 & +5.4 & \textbf{+12.2} \\
\quad Best-of-methods & 70.8 & 19.6 [15.8, 23.7] & - & - & +7.0 \\
\bottomrule
\end{tabular}
\end{table*}

\begin{figure*}[t]
\centering
\includegraphics[width=0.85\textwidth]{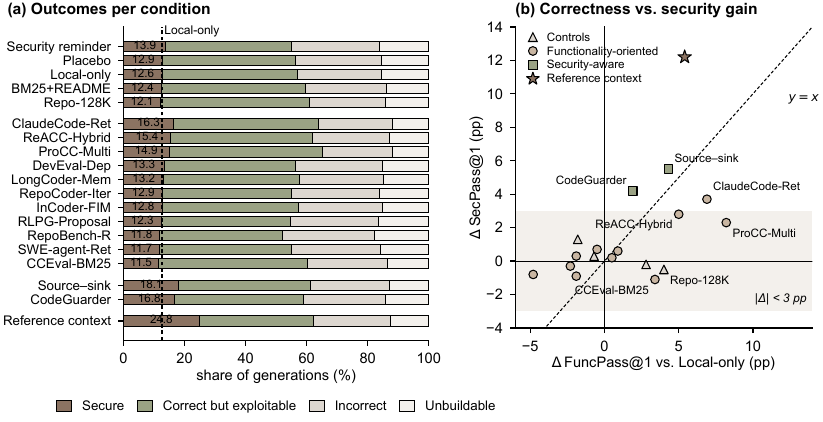}
\caption{RQ1: (a) Outcome partition per condition (number: \SP@1). (b) Change in \FP@1 versus change in \SP@1 relative to Local-only; the shaded band marks changes below the 3~pp practical-relevance threshold.}
\Description{A stacked bar chart shows secure, correct-but-exploitable, incorrect, and unbuildable shares per condition, and a scatter plot compares functional and security gains.}
\label{fig:rq1-partition}
\end{figure*}

\textbf{Functionality-oriented context yields limited security gains.}
Relative to Local-only, the eleven functionality-oriented methods change \FP@1 by $-4.8$ to $+8.2$ pp but \SP@1 by only $-1.1$ to $+3.7$ pp (Table~\ref{tab:rq1-main}, Figure~\ref{fig:rq1-partition}).
Only ClaudeCode-Ret reaches the 3 pp threshold, and no method differs significantly from Local-only after correction (ClaudeCode-Ret: OR 1.35, $p_{\text{Holm}}=.19$); smaller effects cannot be ruled out.
Four methods even have lower observed \SP@1 than Local-only, and 40.3-50.2\% of all generations are correct but exploitable.
This pattern replicates the observation of RealSec-bench~\cite{wang2026realsecbench} for Java retrievers and extends it to all major method families.

\textbf{Correctness-optimal is not security-optimal.}
ProCC-Multi has the highest \FP@1 (65.1\%) but ranks only third in \SP@1, because correct-but-exploitable code accounts for 72\% of its functional gain.
CCEval-BM25 and Repo-128K even raise \CbE@1 by more than \FP@1, so their additional working code is exploitable on balance.
The rankings by \FP@1 and \SP@1 agree only moderately ($\tau_b=0.48$), and a cross-fitted oracle that picks the best method per task exceeds the best single method by 3.3 pp, so the methods are complementary.

\begin{table*}[t]
\centering
\footnotesize
\setlength{\tabcolsep}{3.5pt}
\caption{RQ1: \SP@1 (\%) by CWE family for selected conditions. ``Best func.''\ is the best functionality-oriented method (ClaudeCode-Ret); ``Best-of'' is the cross-fitted oracle.}
\label{tab:rq1-family}
\begin{tabular}{lrrrrrrr}
\toprule
\textbf{Family} & \textbf{Tasks} & \textbf{Local-only} & \textbf{Best func.} & \textbf{Best-of} & \textbf{CodeGuarder} & \textbf{Source-sink} & \textbf{Reference} \\
\midrule
Memory safety & 65 & 16.2 & 21.4 & 25.6 & 21.8 & 19.9 & 26.2 \\
Injection & 42 & 13.4 & 16.8 & 19.8 & 17.7 & 21.2 & 28.0 \\
Access control & 25 & 2.0 & 2.9 & 4.4 & 3.1 & 9.3 & 17.6 \\
Other & 13 & 12.0 & 15.0 & 17.9 & 15.4 & 15.8 & 20.9 \\
\midrule
All & 145 & 12.6 & 16.3 & 19.6 & 16.8 & 18.1 & 24.8 \\
\bottomrule
\end{tabular}
\end{table*}

\textbf{Security-aware context helps, depending on the weakness.}
CodeGuarder adds 4.2 pp (OR 1.40, $p_{\text{Holm}}=.07$) and Source-sink 5.5 pp (OR 1.53, $p_{\text{Holm}}=.006$), and both lower \CbE@1 below Local-only.
Their advantages over the priming controls are not significant after correction, but the Placebo changes \SP@1 by only 0.3 pp, and the Security reminder adds 1.3 pp while lowering \FP@1 by 1.8 pp because some generations reject benign inputs.
The Reference context adds 12.2 pp, 8.5 pp more than the best functionality-oriented method (OR 1.69, $p_{\text{Holm}}=.001$), and Source-sink recovers 4\% of the Local-only-to-Reference gain. 
By CWE family (Table~\ref{tab:rq1-family}), CodeGuarder is strongest on memory safety, where generic checks transfer, whereas Source-sink is strongest on access control (9.3\% vs.\ 3.1\%), where security depends on project policies.
These are method-level comparisons, since the two methods also differ in retrieval logic and context length.
By locality, the best functionality-oriented method gains +2.0, +6.4, and +0.8 pp on L1 to L3 tasks, whereas Source-sink gains +1.8, +6.8, and +7.4 pp and the Reference context +3.8, +14.8, and +17.0 pp.
Similarity retrieval thus occasionally surfaces a project API (L2) but almost never a project policy (L3).
Repo-128K raises \FP@1 by 4.0 pp but lowers \SP@1 by 0.5 pp, so 128K tokens of repository code do not yield more security than a 4K-token slice.

\textbf{Robustness.}
Rankings at native budgets agree with the common-cap ranking ($\tau_b=0.89$), and excluding non-derivable tasks or near-copies leaves all rankings intact.
On mutated tasks, \SP@1 differs by at most 1.6 pp.
With equal project weights, or without the six largest projects, effect sizes change slightly (e.g., Source-sink: $+4.9$ and $+5.2$ pp) but the ranking is preserved ($\tau_b=0.81$).

\begin{finding}{Finding 1 (RQ1): Functionality-oriented context improves correctness more than security.}
Functionality-oriented methods change \FP@1 by up to 8.2 pp but \SP@1 by at most 3.7 pp, and none differs significantly from Local-only. Security-aware methods add 4.2-5.5 pp, with the largest gains of Source-sink on L3 tasks.
\end{finding}

%% ===================================================================
\subsection{RQ2: Which Context Factors Help or Hurt, and Why Does Code Remain Insecure?}
\label{subsec:rq2}

\begin{figure*}[t]
\centering
\includegraphics[width=0.7\textwidth]{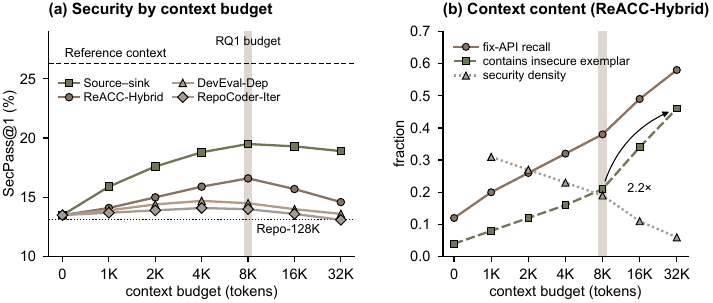}
\caption{RQ2 (amount): (a) \SP@1 by retrieval cap on the three RQ2 models (cap 0: Local-only; shaded: the 8K cap of RQ1). (b) Content of the ReACC-Hybrid context over all 145 tasks.}
\Description{Two panels show secure pass rates by retrieval budget and changes in API recall, insecure exemplar exposure, and security density as the context budget increases.}
\label{fig:rq2-budget}
\end{figure*}

\textbf{Amount.}
\SP@1 of ReACC-Hybrid rises from 13.5\% (Local-only) to 16.6\% at 8K tokens and falls to 14.6\% at 32K (Figure~\ref{fig:rq2-budget}).
Beyond 8K, its fix-API recall still grows, but so does the share of contexts with an insecure exemplar (from 41\% to 90\%), while security density drops from .19 to .06.
The decline thus coincides with more exemplars rather than with lower recall.
DevEval-Dep and RepoCoder-Iter already peak at 4K, Source-sink saturates at 8K because it adds few units beyond the slice, and Repo-128K stays at the Local-only level.

\begin{figure*}[t]
\centering
\includegraphics[width=0.85\textwidth]{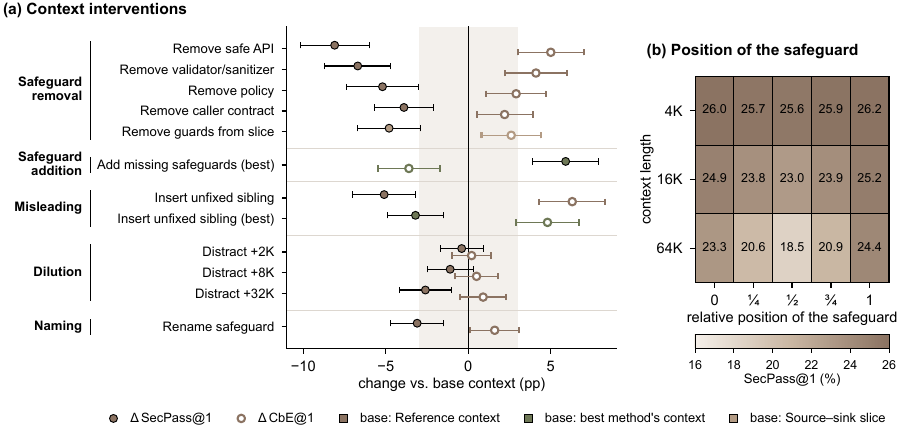}
\caption{RQ2 (content and presentation): (a) Effects of context interventions on \SP@1 (filled) and \CbE@1 (hollow) relative to the unmodified base context on the same tasks and three models, with 95\% intervals; the shaded band marks changes below 3~pp. ``(best)'': applied to the context of the best functionality-oriented method. (b) \SP@1 of the Reference context by the relative position of the safeguard and the total context length.}
\Description{A dot-and-whisker chart compares changes in secure and correct-but-exploitable outcomes after context edits. A heatmap shows secure pass rates by safeguard position and context length.}
\label{fig:rq2-interventions}
\end{figure*}

\textbf{Removing and adding safeguards.}
Removing one safeguard role from the Reference context lowers \SP@1 by 8.1 pp for the safe API, 6.7 pp for validators, 5.2 pp for the policy, and 3.9 pp for the caller contract (Figure~\ref{fig:rq2-interventions}a).
On access-control tasks, removing the policy alone costs 11.3 pp.
Removing the guards from the Source-sink slice shrinks its gain from +6.0 to +1.2 pp, and adding missing safeguards to the best method's context raises \SP@1 by 5.9 pp.
Most lost secure outputs (56-62\%) become correct but exploitable, and the rest become incorrect, so part of the effect is functional.
Replacing a removed role with irrelevant code of the same length at the same position yields nearly the same drops (3.6-7.7 pp), so length and position explain little of the effect.

\textbf{Insecure exemplars.}
On the 74 tasks with an exemplar, inserting one exemplar into the Reference context raises \CbE@1 by 6.3 pp although the safeguard remains present.
In the context of the best functionality-oriented method, the same insertion raises \CbE@1 by 4.8 pp.
Irrelevant code of the same length at the same position raises it by only 0.6 pp, and appending 2K, 8K, or 32K tokens of irrelevant code raises it by at most 0.9 pp.
What a context shows thus matters far more than how much it shows.

\textbf{Presentation.}
The position effect grows with context length (Figure~\ref{fig:rq2-interventions}b): a safeguard in the middle costs 0.6 pp at 4K tokens but 5.9 pp at 64K, mirroring the lost-in-the-middle effect~\cite{liu2024lost}.
Renaming the safeguard costs 3.1 pp, more on web tasks (4.6 pp) than on memory-safety tasks (1.2 pp), which suggests that models partly recognize safeguards by name (e.g., \textit{check\_permission}).

\begin{finding}{Finding 2 (RQ2): Withheld safeguards and inserted exemplars reduce security.}
Withholding a safeguard role costs 3.9-8.1 pp of \SP@1, and supplying missing safeguards adds 5.9 pp. One inserted insecure exemplar raises \CbE@1 by 6.3 pp despite the safeguard, whereas length-matched irrelevant code raises it by 0.6 pp.
\end{finding}

\begin{table*}[t]
\centering
\small
\caption{RQ2: Failure patterns of correct-but-exploitable code (21,945 \CbE{} generations from the 19 RQ1 conditions). Labels may co-occur. ``Share of \CbE{}'': percentage of the \CbE{} generations of each group; ``Per 100 gen.'': the same pattern per 100 generations of the group, i.e., the share multiplied by the group's mean \CbE@1 (44.9\% and 42.6\%). Both groups exclude the six controls and use the same tasks and models. Agreement of automatic and manual labels on 200 samples: $\kappa=0.81$.}
\label{tab:rq5-failures}
\begin{tabular}{p{1.65cm}p{4.3cm}rrrrr}
\toprule
 & & \multicolumn{3}{c}{\textbf{Share of \CbE{}}} & \multicolumn{2}{c}{\textbf{Per 100 gen.}} \\
\textbf{Pattern} & \textbf{Definition} & \textbf{All} & \textbf{Func.} & \textbf{Sec.} & \textbf{Func.} & \textbf{Sec.} \\
\midrule
\textsc{Missing} & no acceptable safeguard set (L2/L3) & 44\% & 46\% & 29\% & 20.6 & 12.3 \\
\textsc{Knowledge} & L1 task; generic defense omitted & 13\% & 13\% & 12\% & 5.8 & 5.1 \\
\textsc{Unused} & present, neither called nor replicated & 16\% & 15\% & 28\% & 6.7 & 11.9 \\
\textsc{Misused} & called, yet exploited & 10\% & 9\% & 22\% & 4.0 & 9.4 \\
\textsc{Imitated} & similar to an exemplar in context & 19\% & 20\% & 9\% & 9.0 & 3.8 \\
\textsc{Other} & wrong threat model, partial fix & 6\% & 6\% & 8\% & 2.7 & 3.4 \\
\bottomrule
\end{tabular}
\end{table*}

\textbf{Failure patterns.}
With functionality-oriented context, most correct-but-exploitable code lacks a safeguard (\textsc{Missing}) or imitates an exemplar (\textsc{Imitated}); with security-aware context, unused and misused safeguards become frequent (Table~\ref{tab:rq5-failures}).
The shift holds per 100 generations, so it is not an artifact of the denominator: \textsc{Missing} falls from 20.6 to 12.3, while \textsc{Unused} and \textsc{Misused} together rise from 10.7 to 21.3.
It also appears within task-model pairs, where the dominant label moves from \textsc{Missing} to a utilization failure in 29\% of pairs and in the reverse direction in 5\%.
Label precision ranges from .72 (\textsc{Imitated}) to .94 (\textsc{Missing}).
In multi-location tasks, 14\% of correct-but-exploitable generations guard the masked region locally but consume an unchecked value from an unfixed sibling, and open coding revealed two further patterns (e.g., defending against the wrong attack).

\begin{finding}{Finding 3 (RQ2): Remaining failures concentrate on applying retrieved safeguards.}
With security-aware context, unused and misused safeguards rise from 10.7 to 21.3 per 100 generations, and the shift also appears within task-model pairs.
\end{finding}

%% ===================================================================
\subsection{RQ3: Do Coding Agents Gather Security-Relevant Context?}
\label{subsec:rq3}

\begin{table*}[t]
\centering
\small
\setlength{\tabcolsep}{2pt}
\caption{RQ3: End-to-end agents versus the best retrieval-based generation method with the same backbone (145 tasks, $n=3$). ``Best'': the functionality-oriented method with the highest RQ1 \SP@1 for that generator. ``SG obs.'': share of trajectories (or contexts) that contain a safeguard; ``IE obs.'': the same for an insecure exemplar, on the 74 tasks that contain one; ``Tokens'': repository tokens observed.}
\label{tab:rq3-agents}
\begin{tabular}{llrrrrrrr}
\toprule
\textbf{Agent} & \textbf{Backbone} & \textbf{\FP@1} & \textbf{\SP@1} & \textbf{\CbE@1} & \textbf{SG obs.} & \textbf{IE obs.} & \textbf{Tokens} & \textbf{Cost} \\
\midrule
SWE-agent & Sonnet 4.5 & 69.7 & 16.6 & 53.1 & 29\% & 46\% & 141K & \$1.08 \\
SWE-agent & GPT-5.4 & 71.0 & 16.1 & 54.9 & 27\% & 43\% & 127K & \$0.91 \\
Claude Code & Sonnet 4.5 & 74.0 & 18.4 & 55.6 & 36\% & 51\% & 204K & \$1.41 \\
\midrule
ClaudeCode-Ret (best) & Sonnet 4.5 & 66.9 & 17.9 & 49.0 & 22\% & 36\% & 7.2K & \$0.09 \\
ReACC-Hybrid (best) & GPT-5.4 & 68.3 & 17.0 & 51.3 & 19\% & 41\% & 7.8K & \$0.05 \\
\bottomrule
\end{tabular}
\end{table*}

\begin{figure*}[t]
\centering
\includegraphics[width=0.8\textwidth]{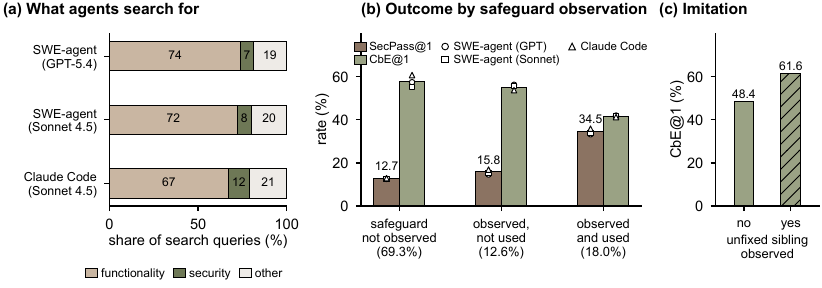}
\caption{RQ3: (a) Share of search queries that name functional entities, security concepts, or other entities. (b) \SP@1 and \CbE@1 of agent trajectories by whether the safeguard was observed and used (bars: pooled; markers: agents; parentheses: share of trajectories). (c) \CbE@1 by whether an insecure exemplar was observed, on the 74 tasks that contain one.}
\Description{Three panels compare agent query categories, outcomes grouped by safeguard observation and use, and correct-but-exploitable rates grouped by insecure exemplar observation.}
\label{fig:rq3-observed}
\end{figure*}

\textbf{Exploration.}
Agents raise \FP@1 over the best retrieval-based method with the same backbone by 2.7-7.1 pp but change \SP@1 by only $-1.3$ to $+0.5$ pp (Table~\ref{tab:rq3-agents}), consistent with prior agent studies~\cite{shen2025secrepobench,zhao2026susvibes,patir2025dualgauge}.
Although agents observe 16-26$\times$ as many repository tokens as an 8K context, they observe the relevant safeguard in only 27-36\% of trajectories, yet an insecure exemplar in 43-51\% of the trajectories on tasks that contain one.
Their searches target functionality: 71\% of queries name functional entities (e.g., a file format), and only 9\% name security concepts (Figure~\ref{fig:rq3-observed}a).
Manual validation agrees with this classification ($\kappa=0.81$), and counting security intent regardless of wording raises the share only to 13\%.

\textbf{Observation and outcome.}
Trajectories that observed and used the safeguard reach 34.5\% \SP@1, versus 12.7\% for those that never observed it (Figure~\ref{fig:rq3-observed}b).
In a mixed-effects model, observing the safeguard is associated with higher odds of a secure outcome (OR 2.4), whereas the scaffold difference is inconclusive (OR 1.2 [0.9, 1.6]).
The association is observational, but in 81\% of these trajectories the safeguard was observed before the first edit, and the association persists for such observations (OR 2.1).
Between 5\% and 11\% of trajectories hit the step or cost cap, and excluding them changes \SP@1 by at most 0.7 pp.
Observation is not sufficient: 41\% of the trajectories that observed a safeguard did not use it and reach only 15.8\% \SP@1.
Conversely, trajectories that observed an insecure exemplar produce more correct-but-exploitable code (61.6\% vs.\ 48.4\%; Figure~\ref{fig:rq3-observed}c).

\begin{finding}{Finding 4 (RQ3): Coding agents mostly search for functionality.}
Agents observe the relevant safeguard in only 27-36\% of trajectories, and most of their queries target functionality. Trajectories that observed the safeguard were more often secure.
\end{finding}

%% ===================================================================
\subsection{Measurement Robustness: Does the Oracle Change the Conclusions?}
\label{subsec:measurement}

\begin{figure}[t]
\centering
\includegraphics[width=\linewidth]{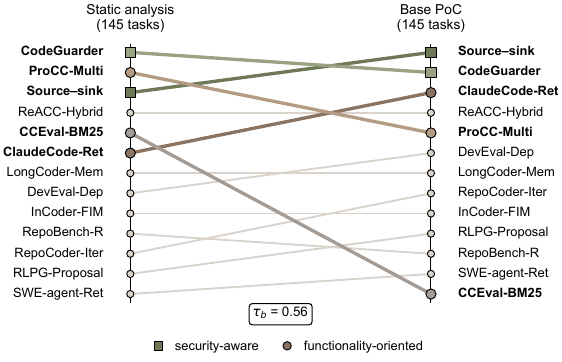}
\caption{Rankings of the 13 methods by security under the static and the base-PoC oracle on the same 145 tasks (top: most secure).}
\Description{A rank comparison links each context method's position under static analysis to its position under the executable exploit oracle.}
\label{fig:rq4-bump}
\end{figure}

\begin{table*}[t]
\centering
\small
\caption{Static analysis versus the PoC oracle on all 28,984 functionally correct RQ1 generations, and the base PoC versus exploit variants on 40 tasks.}
\label{tab:rq4-agreement}
\begin{tabular}{lrr|lr}
\toprule
 & \textbf{PoC fails} & \textbf{PoC succeeds} & \textbf{Summary} & \\
\midrule
\textsc{StaticSecure} & 5,138 & 8,778 & Agreement / $\kappa$ & 63.2\% / 0.25 \\
\textsc{StaticAlarm} & 1,901 & 13,167 & Precision / recall$^\dagger$ & .87 / .60 \\
\midrule
\multicolumn{3}{l}{Resists the base PoC, exploited by a variant (40 tasks)} & Share & 7.3\% \\
\bottomrule
\multicolumn{5}{p{0.94\linewidth}}{\footnotesize $^\dagger$ Static alarms are evaluated relative to the base PoC verdict; PoC-resistant alarms are not necessarily false positives.}
\end{tabular}
\end{table*}

Static analysis and the PoC oracle disagree on 36.8\% of the 28,984 functionally correct RQ1 generations ($\kappa=0.25$; Table~\ref{tab:rq4-agreement}), and the disagreement goes in both directions.
Relative to the PoC verdict, static analysis has a precision of .87 and a recall of .60: it misses 40\% of exploited generations (e.g., off-by-one bounds checks), and it flags 27\% of the generations that resist the PoC, which may be false alarms or vulnerabilities that the PoC does not trigger.
The same pattern appears on code with known verdicts: the queries miss 17\% of the vulnerable versions and flag 13\% of the patched ones.
Exploit variants compromise 7.3\% of the generations that resist the base PoC.
On the same 145 tasks, the two oracles rank the methods only moderately alike ($\tau_b=0.56$; Figure~\ref{fig:rq4-bump}): static analysis favours ProCC-Multi and CodeGuarder, whereas the PoC oracle favours ClaudeCode-Ret and Source-sink.
Comparative conclusions about context methods therefore depend on the oracle, and a static-only evaluation would have favoured methods whose code merely contains defensive patterns.

%% ===================================================================
\subsection{Case Study: A Repository-Specific Policy}
\label{subsec:case-analysis}

\lstdefinestyle{case}{basicstyle=\ttfamily\footnotesize,columns=fullflexible,keepspaces=true,
  frame=single,framerule=0.4pt,rulecolor=\color{gray!60},backgroundcolor=\color{gray!5},
  xleftmargin=3pt,xrightmargin=3pt,aboveskip=2pt,belowskip=2pt,
  showstringspaces=false,breaklines=true,commentstyle=\normalfont\ttfamily,escapeinside={(*}{*)}}

\begin{figure*}[t]
\centering
\begin{minipage}{\linewidth}
\begin{lstlisting}[style=case,language=PHP]
// add_edit_amendments.php, masked region (lines 19-52): patched version, shown for the reader only
(*\textbf{if (!AclMain::aclCheckCore('patients', 'amendment')) \{}*) /* render unauthorized page */ exit; }
// same file, below the masked region: a different ACL object, used only to render the form    (misleading)
$haveAccess = AclMain::aclCheckCore('patients', 'trans');   $onlyRead = ($haveAccess) ? 0 : 1;
// demographics.php: the project's amendment policy                                            (safeguard)
'auth' => AclMain::aclCheckCore('patients', 'amendment', '', 'write'),
// list_amendments.php, print_amendments.php: no check at all                           (unfixed siblings)
\end{lstlisting}
\end{minipage}
\caption{Security-relevant code of CVE-2022-2732 (OpenEMR). The masked region is shown in its patched version for the reader (bold: added by the fix); the model sees the target file without it. All other code is at the base commit.}
\Description{An OpenEMR code excerpt contrasts the required amendment permission with an unrelated form permission and unchecked sibling pages. A generated check placed after the request handler cannot prevent the unauthorized write.}
\label{fig:case-code}
\end{figure*}

We examine CVE-2022-2732 in OpenEMR (CWE-862, L3), the access-control task on which the conditions differ most (18 generations each).
Before the fix, any authenticated user could amend a patient record through \textit{add\_edit\_amendments.php}.
The required permission appears only in a widget guard in \textit{demographics.php} (Figure~\ref{fig:case-code}) and cannot be inferred from the requirement, while the target file and its sibling pages contain misleading or missing checks.
Local-only produced no secure generation, and functionality-oriented methods produced at most three, because similarity retrieval favoured the unchecked siblings (\textsc{Imitated}).
Following CodeGuarder's generic advice, generations checked only for a login, which the PoC user has, or demanded administrator rights, which rejects the legitimate clinician of the functional tests (1/18).
ClaudeCode-Ret retrieved \textit{demographics.php} but ranked it after the siblings (3/18), whereas Source-sink followed the request from its form to the database write and retrieved the guard (6/18).
The Reference context yielded 12 secure generations, but four generations checked the permission only after the request had been processed (\textsc{Misused}).
Retrieving the policy thus did not ensure its correct use on the attack path.
Together with the motivating example, in which similarity retrieval exposed a guard, the case shows that the same method can help on one task and hurt on another, depending on whether similar code carries the safeguard.

\section{Discussion}
\label{sec:discussion}

\subsection{Implications}
\label{subsec:implications}

\textbf{Why functionality-oriented retrieval misses safeguards.}
Our results suggest a common cause.
Functionality-oriented methods build their queries from the target code and the requirement, which describe \emph{what} the code does rather than \emph{which policy} constrains it.
Safeguards, however, often live elsewhere in the repository (e.g., in a central access-control module), so they rarely resemble the target, whereas unfixed siblings resemble it closely.
Agents show the same bias in their search queries (Finding~4), and longer contexts mainly add more siblings (\S\ref{subsec:rq2}).
The motivating example illustrates the exception: similarity helps only when a sibling happens to carry the guard, which makes functionality-oriented retrieval an unreliable source of safeguards.

\textbf{Retrieve safeguards explicitly.}
The removal experiments and the gains of Source-sink support retrieving the project's safeguards (e.g., validators and access policies), especially on L2 and L3 tasks (Findings~1 and~2).
Context budgets should follow safeguard coverage rather than size, because larger contexts also expose more insecure exemplars.
The annotations of \bench{} make such retrievers measurable, since safeguard coverage can be computed for any context before generation.

\textbf{Check examples before reuse.}
An inserted insecure exemplar increases exploitable code even when a safeguard is present (Finding~2).
Since similarity retrieval can surface both secure and insecure siblings (\S\ref{subsec:motivating-example}), retrieved examples should be checked for the required safeguards before they serve as templates.
The static analyses that already run in many pipelines can serve this purpose, because an exemplar only needs to be flagged, not proven insecure.

\textbf{Make agents security-aware.}
Agents rarely search for security concepts, and retrieved safeguards often remain unused or are misapplied (Findings~3 and~4).
These results motivate, but do not validate, two directions: agents that search for the safeguards of a sensitive operation before editing it, and post-generation checks that a retrieved safeguard actually guards the untrusted value.

\textbf{Evaluate context explicitly.}
Because the context changes which methods look best, secure-code benchmarks should treat it as an experimental variable.
They should also report correct-but-exploitable code separately and pair functional tests with executable exploits, since static oracles alone rank methods differently (\S\ref{subsec:measurement}).

\textbf{Future directions.}
Two directions follow from our results.
First, context constructors could identify which kind of safeguard a task needs (e.g., a bounds check or an access policy) before retrieving, since methods are complementary across tasks.
Second, \bench{} can serve as a testbed for such security-aware retrievers and agents, because its annotations allow measuring safeguard coverage before any generation.
Extending it to feature-level tasks that span several files would test whether our findings hold when the edit scope is larger.

\subsection{Threats to Validity}
\label{sec:threats}

\textbf{Construct validity.}
A PoC can demonstrate but not exclude a vulnerability, so \SP@1 is relative to our tests and exploits; exploit variants change 7.3\% of verdicts on 40 tasks, and a manual audit agrees with the oracle on 95.7\% of generations.
The failure labels are heuristic (precision .72-.94), and the static oracle is a single query configuration.

\textbf{Internal validity.}
We compare reproduced context policies, which may deviate from the original systems in newly supported languages.
The agent analysis is observational, and only the removal and insertion edits have matched controls.
Mutated and recent tasks agree with the main results, but contamination cannot be ruled out.

\textbf{External validity.}
The tasks come from 51 projects, and six projects contribute 67 tasks; reweighting projects preserves the pattern, but task characteristics (e.g., weakness type and language) remain partly confounded.
Differences below about 5.6 pp may remain undetected, and the case study illustrates rather than quantifies the mechanisms.

\section{Related Work}
\label{sec:related-work}

\begin{table*}[t]
\centering
\small
\setlength{\tabcolsep}{3pt}
\caption{Repository-level benchmarks for the security of LLM-generated code. ``Func.'': tests check functionality; ``Exploit'': an executable exploit checks security; ``Context'': context is the controlled variable (SecRepoBench and RealSec-bench compare a few retrievers secondarily); ``Annot.'': security-relevant code annotated.}
\label{tab:related-comparison}
\begin{tabular}{lcrcccccc}
\toprule
\textbf{Benchmark} & \textbf{Lang.} & \textbf{Tasks} & \textbf{Real CVEs} & \textbf{Func.} & \textbf{Exploit} & \textbf{Context} & \textbf{Agents} & \textbf{Annot.} \\
\midrule
A.S.E~\cite{lian2026ase} & 5 & 120 & \cmark & \xmark & \xmark & \xmark & \xmark & \xmark \\
SecRepoBench~\cite{shen2025secrepobench} & 2 & 318 & \cmark & \cmark & \cmark & \xmark & \cmark & \xmark \\
RealSec-bench~\cite{wang2026realsecbench} & 1 & 105 & \cmark & \cmark & \xmark & \xmark & \xmark & \xmark \\
SecureVibeBench~\cite{chen2026securevibebench} & 2 & 105 & \cmark & \cmark & \cmark & \xmark & \cmark & \xmark \\
SusVibes~\cite{zhao2026susvibes} & 1 & 200 & \cmark & \cmark & \cmark & \xmark & \cmark & \xmark \\
\midrule
\textbf{\bench{} (ours)} & 7 & 145 & \cmark & \cmark & \cmark & \cmark & \cmark & \cmark \\
\bottomrule
\end{tabular}
\end{table*}

\textbf{Security of LLM-generated code.}
Code assistants often suggest vulnerable code~\cite{pearce2022asleep,perry2023users}.
Secure-code benchmarks have progressed from snippets with static or manual judgment~\cite{siddiq2022securityeval,tony2023llmseceval,bhatt2023cyberseceval,hajipour2024codelmsec,siddiq2024sallm,li2025safegenbench} to executable function-level and application-level tasks~\cite{peng2025cweval,nie2025secodeplt,vero2025baxbench,wang2026codeseceval,chen2026seccodebenchv2,patir2025dualgauge}, and then to repository-level tasks from real vulnerabilities~\cite{lian2026ase,shen2025secrepobench,chen2026securevibebench,zhao2026susvibes,wang2026realsecbench} (Table~\ref{tab:related-comparison}).
A.S.E~\cite{lian2026ase} judges security with static rules.
SecRepoBench~\cite{shen2025secrepobench} and RealSec-bench~\cite{wang2026realsecbench} find that agents and retrievers mainly improve correctness.
Table~\ref{tab:related-comparison} compares \bench{} with these benchmarks along the capabilities that a study of context requires.
Executable functional and exploit oracles are increasingly common, and several benchmarks also evaluate agents.
\bench{} differs from all of them in treating the context as the controlled variable and in annotating security-relevant code, which makes the content of a context measurable.
Security-agent benchmarks~\cite{lee2025secbench,lee2026secbenchpro,mei2024arvo,wang2026cybergym,zhu2025cvebench} and vulnerability repair~\cite{bui2022vul4j,fu2022vulrepair,pearce2023zeroshot} start from a known vulnerability, whereas our models are never told that one exists.
We share their emphasis on executable, effect-based verdicts, and like SEC-bench Pro~\cite{lee2026secbenchpro}, we find that the grading rule can change conclusions.
Our validation and contamination checks follow SWE-bench Verified~\cite{swebenchverified} and studies of evaluation leakage~\cite{matton2024leakage}.

\textbf{Context for repository-level code generation.}
Repository-level benchmarks show that generation depends on cross-file information. Such context~\cite{liu2024repobench,ding2023crosscodeeval,yu2024codereval,jimenez2024swebench} is obtained by the method families of Table~\ref{tab:strategies}, by graph retrieval~\cite{liu2024graphcoder,cheng2024draco,ouyang2025repograph}, or by agents~\cite{zhang2024codeagent,zhang2024autocoderover,xia2025agentless}.
Studies of retrieval~\cite{wang2025coderagbench,gu2025whattoretrieve,hai2025repoexec} and long contexts~\cite{shi2023distracted,liu2024lost,hsieh2024ruler} show that context can also hurt correctness, but they measure only whether code works.
We show how these effects bear on security and identify a mechanism that correctness studies cannot observe: the imitation of insecure exemplars.

\textbf{Improving the security of generated code.}
Model-side techniques (e.g., security hardening~\cite{he2023sven,he2024safecoder} or constrained decoding~\cite{fu2024constrained}) and prompting~\cite{tony2025prompting} change the model or the instruction.
Retrieval-based approaches add \emph{external} security knowledge~\cite{zhang2024seccoder,lin2025securitycourse,shi2026rescue} or show that poisoned corpora induce vulnerable code~\cite{lin2025poisoning}.
We instead study the context that the repository itself provides, which is benign but can still mislead, and we include CodeGuarder as a representative of external knowledge.

\section{Conclusion}
\label{sec:conclusion}

In this paper, we constructed \bench{}, an executable and annotated benchmark for studying how repository context affects the security of LLM-generated code.
Using \bench{}, we conducted a controlled study of 13 context construction methods with six LLMs, together with single-element context edits and an analysis of coding-agent trajectories.
We revealed that functionality-oriented context improves correctness more than security, that withheld safeguards and inserted insecure exemplars reduce security, and that coding agents rarely search for security-relevant code.
Based on these findings, we provided guidelines for retrieving project safeguards and checking retrieved examples.
We believe that our work lays the groundwork for security-aware context construction in LLM coding assistants.

\bibliographystyle{ACM-Reference-Format}
\bibliography{refs}

\end{document}